\documentstyle[prl,aps,preprint,floats]{revtex}
\tightenlines

\def\'{^{\prime}}
\def\avrg#1{{\langle #1 \rangle}}
\def\hq{{\hat q}}

\def\eg{{\it e.g., }}
\def\ie{{\it i.e., }}
\def\etal{{\it et al. }}

\def\etc{{\it etc. }}

\def\half{{\textstyle{1\over2}}}

\def\spose#1{\hbox to 0pt{#1\hss}}
\def\lta{\mathrel{\spose{\lower 3pt\hbox{$\mathchar"218$}}
     \raise 2.0pt\hbox{$\mathchar"13C$}}}
\def\gta{\mathrel{\spose{\lower 3pt\hbox{$\mathchar"218$}}
     \raise 2.0pt\hbox{$\mathchar"13E$}}}

\def\re{ \\ }

\begin{document}
\draft
\title{CMB Broad-Band Power Spectrum Estimation}
\author{
J. Richard Bond,$^{1,2}$}
\address{
$^{(1)}$  Canadian Institute for Theoretical Astrophysics,
University of Toronto,
Toronto, Ontario, Canada
M5S 1A7  \\
$^{(2)}$  Institut d'Astrophysique de Paris, 98 bis, Bd. Arago,
Paris 75014, France}
\maketitle
\begin{abstract}
The natural outcome of theoretical calculations of microwave
background anisotropy is the angular power spectrum ${\cal C}_\ell$ as
a function of multipole number $\ell$. Experimental ${\cal C}_\ell$'s
are needed for direct comparison. Estimation procedures using
statistics linear in the pixel amplitudes as well as the conventional
but less useful quadratic combinations are described.  For most
current experiments, a single broad-band power amplitude is all that
one can get with accuracy. Results are given for the Capri-meeting
detections. Mapping experiments, sensitive to many base-lines, can
also give spectral ``colour'' information, either with a series of
contiguous narrow-band powers or as parameterized by a local
``colour'' index $n_{\Delta T}$ (scale invariant is -2, white noise is
0).  Bayesian analyses of the full first year DMR and FIRS maps give
very similar band-powers (\eg $Q_{rms,PS}=17.9 \pm 2.9 \mu K$ {\it
c.f.} $18.6 \pm 4.7 \mu K$ for $n_{\Delta T}=-2$) and colour indices
(with 1 and 2 sigma error bars) $n_{\Delta T}+3
=2.0^{+0.4;+0.7}_{-0.4;-1.0}$ and $ 1.8^{+0.6;+0.9}_{-0.8;-1.3}$ ({\it
c.f.} the value 1.15 for a ``standard'' scale invariant CDM
model). The 53 and 90 GHz DMR maps, as well as the FIRS map, have
residual short-distance noise
which steepens $n_{\Delta T}$. This residual has so far been modelled by
allowing the pixel error bars to increase, absorbing much of the
effect, but further exploration is needed to see if a second residual
evident in the data --- which is, in part, responsible for the high
$n_{\Delta T}$ --- is from systematic errors or is physical.
\end{abstract}

\vspace{.2in}
\noindent
CITA-94-5

\noindent

\newpage
``Un-cleaned'' data with inadequate ``dirt'' models do not yet give us
confidence that what is primary anisotropy has been well-separated
from secondary backgrounds, foregrounds and instrumental systematics
in {\it any} experiment. But we will. In the meantime, we need a
phenomenology to display experimental results that (1) allows us to be
unafraid of presenting data with the inevitable residual contaminants,
(2) gives a $\Delta T/T$ estimator which is not sensitive to the
specific experimental configuration (thus not {\it rms} anisotropies),
(3) allows a meaningful comparison among experiments and (4) among
theories.  The angular $\Delta T/T$ power in broad bands that cover
the multipole range explored by a given experiment satisfies these
requirements. Whether the data allows one or many bands to be
well-estimated depends upon the details of the experiment.

Gaussian theories are completely
characterized by the power spectrum,
${\cal C}_{T\ell} \equiv \ell (\ell +1) \avrg{  \vert a_{\ell
m}\vert^2 }
 /(2\pi)$, where ${\Delta T \over T_0 } (\hq ) =
\sum_{\ell m} a_{\ell m}Y_{\ell m}
(\hq )$ defines the $a_{\ell m} $.
The signal $({\Delta T/ T})_p$ from the $p^{th}$ pixel of
a CMB anisotropy experiment can be expressed in terms of
linear filters  ${\cal F}_{p,\ell m
}$ acting on the $a_{\ell m}$: $\big(\Delta T /T \big)_p =
\sum_{lm} {\cal F}_{p,\ell m }
a_{\ell m}$.
The associated pixel-pixel correlation function
can be expressed in terms of a quadratic $N_{pix}\times
N_{pix}$ filter matrix
$W_{p p^\prime , \ell}$ acting on ${\cal
C}_{T\ell}$, while its trace defines the average filter
$\overline{W}_{\ell}$ (\eg Bond
1989):
\begin{eqnarray}
C_{Tpp^\prime} &\equiv &
\avrg{\Big({\Delta T \over T}\Big)_p\Big({\Delta T \over
T}\Big)_{p^\prime}}  = {\cal I}\Big[ W_{p p^\prime , \ell}{\cal
C}_{T\ell} \Big] \ , \quad  W_{p p^\prime , \ell} \equiv {4\pi \over
2\ell +1}  \sum_{lm} {\cal F}_{p,\ell m } {\cal F}_{p^\prime ,\ell m
}^* \ ;
\label{eq:pspec.ctpp} \\
\Big({\Delta T \over T}\Big)^2_{rms} &\equiv &
{1\over N_{pix} } \sum_{p=1}^{N_{pix}} C_{Tpp} \equiv
   {\cal I}\Big[ \overline{W}_{\ell}{\cal
C}_{T\ell} \Big] \ , \quad  \overline{W}_{\ell} \equiv {1\over N_{pix} }
\sum_{p=1}^{N_{pix}}  W_{p p , \ell} \ ; \label{eq:pspec.wbar} \\
{\cal I} (f) &\equiv & \sum_\ell f_\ell {(\ell +\half \over \ell
(\ell +1) } \ . \label{eq:logI}
\end{eqnarray}
The ``discrete-logarithmic integral'' ${\cal I}$ of a function
function $f_{\ell}$ is defined by eq.~\ref{eq:logI}. Although there
are as many as $N_{pix}(N_{pix}+1)/2$ different filters to probe ${\cal
C}_{T\ell}$, in practice symmetries reduce this number to at most
$\sim N_{pix}$ and often only a few are large enough to be effective
probes.

\subsubsection{Broad-Band Power and Angular Colour Measures }
\label{Bpow}

If the power spectrum changes slowly over the band that
$B_\ell= \overline{W}_{\ell}$ stretches over, one estimate of the broad-band
power is just a renormalization of the experimental {\it rms}:
\begin{equation}
\avrg{{\cal C}_\ell }_{B} = (\Delta T /
T)^2_{rms}/ {\cal I}[ \overline{W}_{\ell}] =
 {\cal I}[ \overline{W}_{\ell}{\cal
C}_\ell ]/ {\cal I}[ \overline{W}_{\ell}] \ . \label{eq:pspec.rms}
\end{equation}
For an experiment probing isolated (uncorrelated) pixels, this
broad-band power is {\it all} you can get (since
$C_{Tpp^\prime} = C_{T11}\delta_{pp^\prime}$).  Current
intermediate and small angle experiments have insensitive $W_{p p^\prime ,
\ell}$ for pixel separations beyond the first few (\ie
$C_{Tpp^\prime}$ approaches zero rapidly): splitting the
$\overline{W}$-band into $B_1,B_2,...$
sub-bands does little good. However, for
mapping experiments like DMR and FIRS the number of pixel base-lines
probed (number of useful $W_{pp^\prime ,\ell}$) is large so
the large $\overline{W}$-band can be split
into a number of contiguous sub-bands (but not too many
or the error bar matrix becomes too large and complicated). ``Spectral
colours'' can be defined as logarithmic differences of the band-powers.
Alternatively one can assume a power spectrum with a local colour index
$n_{\Delta T}$ as well as a broad-band power amplitude $\avrg{{\cal
C}_\ell}_{B}$:
\begin{eqnarray}
{\cal C}_{B\ell} &=& \avrg{{\cal C}_\ell }_{B} \ (\ell
+\half)^{2+n_{\Delta T}}\ {\cal U}_{B\ell} \
\ {\cal I}[ \overline{W}_{\ell}]/
{\cal I}[ \overline{W}_{\ell} (\ell
+\half )^{2+n_{\Delta T}} {\cal U}_{B\ell}  ] \ . \label{eq:plaw}
\end{eqnarray}
Here an extra input estimator shape ${\cal U}_{B\ell}$ is included to give more
flexibility for statistical testing.  I usually
choose ${\cal U}_{B\ell} =1$, although I find it is always instructive
to see what a matched form, ${\cal U}_{B\ell} \propto
\overline{W}_\ell$, gives. As we learn more experimentally about the
true shape, ${\cal U}_{B\ell}$ can be matched to it.

The amplitudes $\{[\avrg{{\cal C}_\ell}_{B_1}]^{\half},[\avrg{{\cal
C}_\ell}_{B_2}]^{\half},... \} $ can be determined by whatever
statistical method we are most enamoured with, whether Bayesian or
frequentist. Their likelihood function defines a
multidimensional surface, whose peak gives the best band-power
estimates and whose curvature matrix about the peak defines the error
bar matrix; or, better, Bayesian credible regions can be defined for error
estimation.

It also turns out to be simple to transform old-style presentations of
$\Delta T/T$ detections and limits to the band-power and colour
language.

The infamous $(\Delta T /T)_c$-curves as a function
of coherence angle $\theta_c$ can be translated into the more
useful band-power estimates by noting the power spectrum for a
``Gaussian correlation function'' is of the form eq.~\ref{eq:plaw}:
\begin{eqnarray}
{\rm GCF:}\  n_{\Delta T}&=& 0 \ \& \
{\cal U}_{B\ell}=e^{-u^2/2} \ , \ u\equiv
(\ell+\half)/(\ell_c+\half) \ , \ \ell_c+\half \equiv
[2\sin(\theta_c /2)]^{-1} \ ; \label{eq:gcf} \\
\avrg{{\cal C}_\ell }_{B} &\approx & (\Delta T
/T)_c^2 \ {\cal I}\Big[u^2 e^{-u^2/2}  \overline{W}_{\ell} \Big]
 /  {\cal
I}[ \overline{W}_{\ell}] \ . \label{eq:gcfpow}
\end{eqnarray}

Another form often adopted is the low-$\ell$ ``Sachs-Wolfe'' power for
scalar metric perturbations (\eg Bond and Efstathiou 1987),
\begin{eqnarray}
{\cal C}_\ell &=& {\cal C}_2 \ {\ell (\ell +1) \over 6} \
{\big( \ell -{3-n_s \over
2}\big)! \over \big( \ell +{3-n_s \over 2}\big)! } \ {\big( 2+{3-n_s
\over 2}\big)! \over \big( 2- {3-n_s \over 2}\big)! } \ ,
\ \ n_{\Delta T} \approx  n_s-3 \ .  \label{eq:swlaw}
\end{eqnarray}
It differs {\it very} little from the more natural form for
phenomenology, eq.~\ref{eq:plaw}, with the index $n_{\Delta
T} $ related to the primordial density fluctuation index $n_s$ as
shown. Thus $n_s=3$ is white noise in $\Delta T$.

For the DMR and FIRS beams (including pixelization effects), the
relation between the quadrupole power ${\cal C}_2$
(or equivalently $Q_{rms,PS}^2$) and the band-power is
\begin{eqnarray}
\Big({Q_{rms,PS} \over T_0}\Big)^2 &\equiv & {5\over 12 } \, {\cal
C}_2 \approx
{5\over 12 } \, \avrg{{\cal C}_\ell }_{B}\ e^{-\alpha
(n_s-1)(1+0.3(n_s-1))}\ , \ \ \alpha_{dmr}=0.73 \ ,\
\alpha_{firs}=1.1  \ , \label{eq:qrmsps}
\end{eqnarray}
where $T_0 =2.726 \pm 0.01 \,
K$. An advantage of the band-power over $Q_{rms,PS}^2$ is that the former
is roughly independent of $n_{\Delta T}$, while the latter is quite
sensitive to it.

\marginpar{Fig. 1}

Fig.~1 shows estimates for the power spectrum averaged over bands in
$\ell$-space, assuming only wavelength-independent Gaussian
anisotropies in $\Delta T/T$ are contributing to the (sometimes
cleaned) signals. See the reference list for the experiments'
acronyms. In Fig~1(a), the data points denote the maximum
likelihood values for the band-power and the error bars give the 16\%
and 84\% Bayesian probability values (corresponding to $\pm 1\sigma$
if the probability distributions were Gaussian). The horizontal
location is at the average value $\avrg{\ell} \equiv {\cal I} (\ell
\overline{W}) /{\cal I} (\overline{W})$; except for DMR,
FIRS and Python, $\avrg{\ell}$ is near the
$\overline{W}_{\ell}$-maximum.  The horizontal error bars denote where
the filters have fallen to $e^{-0.5}$ of the maximum.

Proceeding from small $\ell$, {\it Qdmr} is the quadrupole power
estimated using the 53GHz maps with a Galactic cut $b_{Gcut}=20^\circ$
(Bond 1993). The treatment of the {\it dmr}, {\it firs} and Tenerife
broad-band powers are described later.  The {\it sp91} band-estimates are
for the 9 point scan, the 13 point scan and for the combined
statistical analysis of the $9+13$ point scan (higher value).
The offsets are for
clarity. All 4 channels were simultaneously analyzed (Bond 1993).
(The $4th$ channel of the 9 point scan gives a $1.5 \times 10^{-10}$ 95\%
credible upper limit.  Including a GCF with a synchrotron slope opens
up the error bars of SP91 considerably.)
The two recent MAX ({\it M})
results are for the scans in
GUM (upper) and Mu Pegasus (lower, with a strong
dust signal removed). The dotted lines ending in triangles denote the
quoted 90\% confidence interval for the MSAM single
({\it g2}) and double ({\it g3})
difference configurations for that half of their data that did not
have obvious large sources in it.
The dashed error bar shown next is the 84\% upper credible
limit for the SP89 9 point scan.  The OVRO 7 point upper limit
is last.
For most experiments with detections, the 50\% Bayesian probability
(where the vertical and horizontal error bars cross)
lies very close to the likelihood maximum.

Fig.~1(b) gives band-powers for other experiments reported at the
Capri meeting that I have estimated from
GCFs (evaluated at or near the most sensitive angle
$\theta_c$) or $(\Delta T/T)_{rms}$,
but that I have not yet done a full
Bayesian analysis for. The BigPlate result is denoted by {\it bp},
Python ({\it py}) is next, followed by Argo.
These higher MSAM2 and MSAM3 results (than in Fig.~1(a))
are what one
gets when the ``source-half'' of the data is also included.
Finally the upper limit from the $m=2$ mode analysis of the
WhiteDish experiment is {\it wd2}.   The $dmr_2$ number --- with the
smaller error bar than the $dmr$ one
in Fig.~1(a) --- is a translation of the
Bennett \etal (1994) result for the second year DMR data,
$Q_{rms,PS}=17.6\pm 1.7 \mu K$ for $n_s=1$, to a
band-power. Given my experience that
transformed band-powers are in excellent agreement
with the results of full Bayesian analysis, I am
confident that Fig.~1(b) gives a fair account of the implications of
the new detections reported at the Capri meeting.

To show how well the transformation formulae
Eqs.~(\ref{eq:pspec.rms}, \ref{eq:gcfpow}, \ref{eq:qrmsps}) work, I use
some of the numbers quoted for the double difference Tenerife
experiment ($5.6^\circ$ beam and $8.2^\circ$ throw)
at the Capri meeting. For the 15 and 33 GHz data,
we heard that $Q_{rms,PS}=26\pm 6
\mu K$ for $n_s=1$, hence using (\ref{eq:swlaw}) gives  $\avrg{{\cal C}_\ell
}_{B} = 2.2^{+1.1}_{-0.9} \times 10^{-10}$. They also
give $(\Delta T)_c=54^{+14}_{-10} \mu K$ at $\theta_c =4^\circ$
($\ell_c=13.8$). The ratio
${\cal I}\Big[u^2 e^{-u^2/2}  \overline{W}_{\ell} \Big]
 /  {\cal
I}[ \overline{W}_{\ell}]$ turns out to be 0.587, hence the
transformation (\ref{eq:gcfpow}) gives $\avrg{{\cal C}_\ell
}_{B} = 2.3^{+1.3}_{-0.8} \times 10^{-10}$. For
all the 15 GHz data (covering $700\, {\rm deg}^2$),
Rebolo gave $(\Delta T/T)_{rms}=39 \pm 10 \mu K$. I find ${\cal
I}[ \overline{W}_{\ell}] =0.98$, hence  $\avrg{{\cal C}_\ell
}_{B} = 2.1^{+1.2}_{-0.9} \times 10^{-10}$ using (\ref{eq:pspec.rms}).
Thus, all are quite
consistent.

Overlaying band-powers on theoretical ${\cal C}_\ell$ curves
basically shows the status of the theories in question. However, since we do
not know precisely where the observed power $\avrg{{\cal C}_\ell
}_{B,obs}$ lies in the band, the direct comparison should be
with the
theoretical band-power $\avrg{{\cal C}_\ell }_{B,th}={\cal I}[{\cal C}
\overline{W}]/{\cal I}[ \overline{W}]$.  These  are shown in Fig~1(c)
for a few representative theories to illustrate the precision of
power spectrum determination we need to finely differentiate among
models. Error bars of 10\%, the best DMR can
possibly do, and that are not
over-optimistic for intermediate- and small-angle mapping experiments,
are also shown. To consider what is required for this accuracy,
consider an experiment with $N_{pix}$ pixels with per-pixel error
$\sigma_{pix}$. Suppose the pixels are sufficiently separated that
only $\overline{W}_\ell$ is an effective probe of ${\cal C}_\ell$.
For large $N_{pix}$, the $\nu$-sigma uncertainty in the experimental
value of the band-power is
\begin{eqnarray}
\avrg{{\cal C}_\ell }_{B,obs} &= &
\avrg{{\cal C}_\ell }_{B,maxL}
  \pm \nu \sqrt{2/N_{pix}}\,  \,
\Big[\avrg{{\cal C}_\ell }_{B,maxL} +
    \sigma_{pix}^2/ {\cal I}[\overline{W}_\ell ]\Big] \, ; \label{eq:powvar} \\
 \avrg{{\cal C}_\ell }_{B,maxL} &= &
\avrg{{\cal C}_\ell }_{B,th}  \pm \nu \sqrt{1/N_{pix}}\,
\Big[\avrg{{\cal C}_\ell }_{B,th} +
    \sigma_{pix}^2/ {\cal I}[\overline{W}_\ell ]\Big] \   . \label{eq:maxvar}
\end{eqnarray}
To get 10\% error bars requires experimental noise to be small {\it
c.f.} the signal (we are basically already there), and $N_{pix} =200$,
\ie a mapping experiment.  For large $N_{pix}$, the observed
maximum likelihood will fluctuate from $\avrg{{\cal C}_\ell }_{B,th}$,
the quantity we want, according to eq.~\ref{eq:maxvar}, but the
error bars of eq.~\ref{eq:powvar} include these
realization-to-realization fluctuations (thus $\sqrt{2}$ appears, not $1$).

\subsubsection{The DMR and FIRS Colour Problem}\label{dmrcolour}

In this section, I describe some Bayesian results on the FIRS and
first year DMR maps. The methods deal with all aspects of the maps
simultaneously and so are highly sensitive to all components in it,
whether they are the primary signals we are interested in or the
warts that Dave Wilkinson admonishes us to be aware of and beware of,
for they blemish ``the face'' that George Smoot (and we too) wish to
contemplate. As we shall see, it seems to be pimples and not warts
which I have to concentrate on cleaning up. Although I think even now
I am doing this moderately well --- for within the statistical observing
strategies I use certain natural pimple-erasers are suggested --- I
still believe that the acne pattern colours my high precision look at the
face. For the issue is one of colour.

To make the DMR map data tractable for
exploration, I used one lower resolution scale than the original maps,
\ie $5.2^\circ$ rather than $2.6^\circ$ pixels.  The dipole and average
subtractions were done after the Galactic latitude cut (taken here to
be $\vert b\vert > 25^\circ$), but before the lowering of the
resolution. I used the Kneissl and Smoot (1993) revision
of the DMR beam (which is non-Gaussian),
and included corrections for digitization and pixelization, especially
important in view of the $5.2^\circ$ pixel size.  Since COBE actually
measures the difference between $\Delta T$ values at 2 beam-smeared
points $60^\circ$ apart, the pixel errors are correlated.  A
correction linear in the off-diagonal terms similar to one proposed by
Lineweaver and Smoot (1993) has been applied to take this effect into
account. The unknown amplitude in the average and quadrupole of the
theory have been handled by treating them as marginal variables to be
integrated over in the Bayesian analysis.

The FIRS (`MIT') map ($168$ GHz) has a highly inhomogeneous weighting
of each of the $1.3^\circ$ pixels. For the results shown here one
lower resolution level was used, with $2.6^\circ$ pixels, although I
have checked that $1.3^\circ$ pixels gives the same answer for the
$n_{\Delta T}=-2$ case. The average and dipole were
removed in the map construction.  To include the effects of
pixelization and the best beam-smearing estimate, a $4.2^\circ$ beam
was used rather than $3.8^\circ$. (The differences in the amplitudes
obtained are quite small.)

A full Bayesian analysis of maps requires frequent inversion and
determinant evaluations of $N_{pix}\times N_{pix}$ correlation
matrices, the sum of all $C_{Tpp^\prime}$ in the theoretical modelling
plus the pixel-pixel observational error matrix $C_{Dpp^\prime}$.  I
transform this matrix and the amplitude vector $\vec{\Delta}
 = (\Delta_p)$ to ones
that absorb the pixel 'weighting', ${\cal E}_T \equiv C_D^{-1/2} C_T
C_D^{-1/2} $ and $\vec{x}\equiv C_D^{-1/2} \vec{\Delta}$, with `dimensions'
of $(signal-to-noise)^{2,1}$, respectively.  The removal of averages, dipoles,
gradients, \etc are included as more `theoretical' signals, whose
`unobserved' contributions to $\Delta_p$ are integrated over, creating
marginal likelihood functions. For their prior distribution, very
broad Gaussians are adopted (essentially the same as uniform priors,
but they regularize the inversions). I rotate to the eigenvectors of
${\cal E}_T$, which are fully orthogonal (uncorrelated)
signal-to-noise modes for the map, with eigenvalues $1+{\cal
E}_{TR,k}$ and signals $\xi_k$ for each mode $k$, where ${\cal
E}_{TR}\equiv R{\cal E}_TR^{tr}$ and $\vec{\xi} \equiv R\vec{x}$.  With
uniform weighting and all-sky coverage, these modes $k$ are just the
independent $\Re{(a_{\ell m})}$ and $\Im{(a_{\ell m})}$ for any theory, but
with Galactic cuts followed by dipole removals they are complicated
and theory-dependent.  Even so, these modes are quite instructive
linear combinations of the pixels. Often only a fraction of the modes
are highly sensitive to theories being tested. For example, pixel
differences within the beam are theory-insensitive, but highly
susceptible to excess noise in the experiment, and the
($S/N$)-eigenmodes can be used for effective filtering
or weighting strategies.

I sort the ($S/N$)-eigenvalues, ${\cal E}_{TR,k}$, in decreasing
values. These are plotted as open circles
in Fig.~2 for DMR and FIRS, for the cases
$n_s=1$ {\it (a,b)} and $n_s=2$ {\it (c,d)};
the 1-sigma cosmic variance are the
two solid lines surrounding the points.  ${\cal E}_{TR,k}$ is
normalized to $\sigma_{th}=1$, where $\sigma_{th}^2 \equiv
\avrg{{\cal C}_\ell}_B/10^{-10}$; this just happens to be the
amplitude that the data prefers. Although it may appear from the
${\cal E}_{TR,k}$ decline that only the first few modes are effective
at probing the theory in question, for uniform weighting and all-sky
coverage it is basically equivalent to plotting ${\cal C}_\ell /(\ell
(\ell +1))$, which falls with $\ell$ even without beam-smearing
effects.  A rough way to correct the visual impression is to multiply
by $k$ (which would be $(\ell +1)^2$ for homogeneous-weighting and
all-sky coverage): $k{\cal E}_{TR,k}$ is then nearly constant,
decaying mostly because of the beam.

The reason I didn't plot it this way is to make evident the
(almost flat) excess power at high $k$ in the data,
which the beam-filtered theoretical points
cannot possibly explain. The observational (solid) points in Fig.~2
are the average ($S/N$)-mode band-powers $\beta$,
\begin{equation}
\overline{{\cal P}}_{S/N, \beta}\equiv {1\over n_\beta}
\sum_{k\in \beta} (\overline{\xi}_k^2 -1) \ ,
\quad n_\beta \equiv \sum_{k\in \beta}
1\  . \label{eq:snbpow}
\end{equation}
The (sparse)-binning was chosen to make the data trends clear.  The
error bars are estimated directly from the data, and include a
correction for large scale power as well as a pixel noise error.

The form of Fig.~2 was chosen to show a pictorial procedure for
fitting the data: $\sigma_{th}$ is adjusted until the error envelope
on $\sigma_{th}^2 \, {\cal E}_{TR,k}$, which also scales with
$\sigma_{th}^2$, encompasses the data points as well as
possible. Clearly it will not do very well for the high $k$-bins for
either the $n_{\Delta T} =-2$ or $-1$ theory.  A first approximation
to the residual `noise', which is quite good for the high $k$ end of
the FIRS data, is to assume a constant offset.  This can be viewed as
increasing the error bars by a factor $(1+r)$ (so the offset is
$r^2+2r$). Equivalently I can parameterize this residual noise by a
new `theory' source to be added to $\sigma_{th}^2 \, {\cal E}_{TR,k}$,
with $(S/N)$-power $\sigma_{res}^2 \overline{{\cal E}_{TR}}
\equiv r^2+2r$, where $\overline{{\cal E}_{TR}}\equiv N_{pix}^{-1}
\sum_k {\cal E}_{TR,k}$, arranged so that if $\sigma_{res} =
\sigma_{th}$, the power would be equal.

Of course, this pictorial procedure with quadratic combinations of the
modes is not the best statistical approach to deal with the data. Once
one has transformed to the $(S/N)$-eigenmode system, a Bayesian
analysis on the full map is very straightforward and fast, even with
average and gradient subtractions. The effect of
these subtractions is hard to show in Fig.~2: the $k$-modes are no
longer statistically independent.
(The average affects the first bin, but the dipole has
influence out to $k \sim 30$, and is responsible for some of the power
loss. Note that the observational $\beta$-powers can become negative.)
The first step in the Bayesian method is the construction of a
joint likelihood function in $\sigma_{th}$ and $\sigma_{res}$.
The contour  maps in Fig.~3
show a strong maximum in $\sigma_{th}$ at almost the same level for
the DMR 53A+B, 90A+B, and FIRS maps, in spite of the large differences
in the position of the maximum in $\sigma_{res}$.

The single broad-band powers for DMR and FIRS plotted in Figure 1 are
derived from the marginal distributions in signal amplitude, found
after integrating the joint distribution over all possible residual
noise amplitudes (\ie integrating over the $\sigma_{res}$ axis of
Fig.~3).  For $n_s=1$, $Q_{rms,PS}\equiv 17.6 \mu K \ (\avrg{{\cal
C_\ell}}_B/10^{-10})^2$. I shall express the results in the
$Q_{rms,PS}$ terms: for the $53A+B$ GHz map, I get $17.9 \pm 2.9
\mu K$ ($17.6 \pm 2.8 \mu K$ if I ignore off-diagonal terms in
$C_{Dpp^\prime}$, $17.2 \pm 2.5 \mu K$ if I use a $7^\circ$ Gaussian
beam with no pixelization correction). This compares with the DMR
team's best estimation for the first year data of $17.1 \pm 2.9 \mu K$
(Wright \etal 1994) and the new two-year result of $17.6 \pm 1.7 \mu
K$. For 90A+B, I get $18.6 \pm 3.8 \mu K$ and for 31A+B, $15.8 \pm 5.8
\mu K$.  For 53A-B, I get $Q_{rms,PS}=5.3^{+5.1}_{-5.3} \mu K$, while for
90A-B, $0.0^{+5.1}_{-0.0} \mu K$: \ie no spurious large scale power in the
difference maps. (Curiously I find no clear detection in 90A;
however the error bars for this map are quite
large.)  For $n_s=1$ and the FIRS map, the level is $18.6^{+4.8}_{-4.6}
\mu K$ -- \ie at the DMR level.

The probability for the residual noise level is found by integrating
over the $\sigma_{th}$-axis. The most probable value is indicated by
the horizontal dotted line joining the triangular points in Fig.~2,
about at the level the eye would pick.  For FIRS, zero in residual
noise is excluded at more than the 10 sigma level (with $r=0.25$ most
probable, independent of $n_{\Delta T}$
(Fig.~2(c) {\it c.f.} Fig.~2(a)).  However, the high plateau around
$k=200$ in Fig.~2(a,c) suggests $\sigma_{pix}$-enhancement is not the
whole story (in accord with failure by the FIRS team to identify the
source of the residual after exhaustive checks).

For the first year DMR data, the residual depends upon the map and
upon the spectral steepness. I find no evidence for a residual in the
31A+B map, but do have them in both 53A+B and 90A+B. The residuals in
53B ($r=0.07$) and 90B ($0.09$) are larger than in 53A ($0.01$) and
90A ($0.02$). 53(A-B) has residuals about the same as 53A+B,
suggesting that it is not the effect of physical sources on the
sky. The numbers in brackets are for $n_{\Delta T}=-2$. They drop with
increasing $n_{\Delta T}$ (see Fig.~2 for $n_{\Delta T}=-1$).

Al Kogut and the DMR team have checked the scatter in their data
before conversion to a map and confirm the trends I find in $r$, but
with lower amplitudes than my $n_{\Delta T} =-2 $ values, more like my
$n_{\Delta T} =-1 $ values, the index the data prefers.  However,
there are effects not usually included in map analysis which will
increase the variance, \eg the ``unobserved'' Galactic plane pushes up
the variance in off-plane pixels through the $60^\circ$
correlation. Modelling the excess in Fig.~2(c,d) by a constant is
clearly not optimal. The shape of the data points is
suggestive of beam-effects not being properly included. However, even
with the old standard $7^\circ$ beam (quite different than the Kneissl
and Smoot 1993 beam), and with no pixelization corrections, the ${\cal
E}_{TR,k}$ curve still falls off faster than the data.

Spectral indices from $n_{\Delta T} =-3$ to $n_{\Delta T} =0$ in steps
of $0.1$ were run
to construct likelihood functions in $\{ n_{\Delta
T},\sigma_{th},\sigma_{res}\}$. Integrating first over $\sigma_{res}$
(marginalizing it) allows one to construct $n_{\Delta
T}$-$\sigma_{th}$ contour maps, not shown here. Integrating again,
over $\sigma_{th}$, gives the probability distribution for $n_{\Delta
T}$. The colour index I get is $n_{\Delta T}+3
=2.0^{+0.4;+0.7}_{-0.4;-1.0}$ for the DMR53A+B map and $
1.8^{+0.6;+0.9}_{-0.8;-1.3}$ for the FIRS map, with the second errors
given denoting 2-sigma values. The reason for such steep $n_{\Delta
T}$ is evident from Fig.~2: higher $n_{\Delta T}$ gives a slower fall
of the mode power with increasing $k$. However, one cannot increase
$n_{\Delta T}$ too much or the flattening of ${\cal E}_{TR,k}$ at low
$k$ becomes too much for the high amplitude data points there to
contend with: $n_{\Delta T}=-1$ is the compromise.

It might be thought that filtering out the high $k$ modes will allow
us to avoid modelling the excess found there. However, so far I find
that sharp $k$-filtering, keeping only a few hundred modes, does not
help for FIRS. The $k \sim 200$ plateau keeps the preferred index
high. For DMR with $k \le (16)^2$, the residual $r$ grows for low
$n_{\Delta T}$, stealing some of the power from $\sigma_{th}$ (which
drops by 20\% for $n_{\Delta T}=-2$).  However, the $n_{\Delta T}$
value (and error bars) remain about the same as when all of the modes
are included. For $k \le (13)^2$, $n_{\Delta T}$ becomes poorly
determined, $1.0^{+1.1;+1.6}_{-0.7;-1.0}$. If a faulty beam structure
is responsible, adding to the likelihood function a beam-smearing
parameter could be used to set it by finding the most probable
value. However, a $7^\circ$ Gaussian beam and no pixelization
corrections gives $1.7^{+0.4;+0.7}_{-0.4;-0.9}$, still steep {\it
c.f.} the inflation prediction.  Models with three or more components
have yet to be explored. A particularly interesting case is a white
noise $n_{\Delta T}=0$ distribution to model small scale
(non-primary?) sources, plus a large scale power source with variable
$n_{\Delta T}$, plus pixel error bar enhancement.

Most people probe the data using quadratic combinations ${\cal Q}_A$
of the pixel amplitudes $(\Delta T/T)_p$, and hence of $\xi_k$:
${\cal Q}_A= \sum_{kk^\prime} \gamma_{kk^\prime}^A \xi_k
\xi_{k^\prime}$. The correlation function and various combinations
for the ${\cal C}_\ell$-spectrum are examples, as, of course, are the
$(S/N)$-powers ${\cal P}_{S/N, \beta}$.  The sum of the ${\cal
P}_{S/N, k}$ linearly weighted by the $(S/N)$-eigenvalues, $\sum_k
{\cal E}_{TR,k} \xi_k^2$, is the Boughn-Cottingham (BC) statistic. Its
effect can be seen in Fig.~2: multiplying the data by ${\cal
E}_{TR,k}$ strongly suppresses the contribution of the deviant high
$k$-modes to the sum. A pictorial exercise such as in Fig.~2, but with
only one data point, can be done for the BC statistic. A better method
is to construct a likelihood using many (signal-plus-noise) Monte
Carlo realizations of the $\xi_k$, with average and dipole
subtraction.  This {\it is} the map. Thus, map construction is very
fast since it involves just $N_{pix}$ random number choices and no
direct summation over $Y_{\ell m}$ at each pixel. This high speed
helps, since to create a reasonably smooth likelihood function (rather
than just using frequentist statistical measures) requires fifty
thousand rather than a few thousand realizations. The eigensystem is
of course also excellent for sets of quadratic statistics, \eg for
angular bins of the correlation function and $\ell$-bins of various
prescriptions for quadratic $\ell$-space power spectrum
estimators. One just has to rotate the pixel-pixel quadratic operators
into the $\gamma_{k k^\prime}^A$ form in $k$-space.

The BC quadratic allows good recovery of the most probable Bayesian
values (but it costs more in computer time since no Monte Carlo runs
are required for the Bayesian calculation). With appropriate
normalization, the BC statistic turns out to be the most sensitive
single quadratic estimator of the band-power if there were no dipole
\etc removals and no other signals in the data.  For FIRS, the BC
value is a little bit high. I have therefore constructed best single
band-power estimators with these effects included. For FIRS, the
$\avrg{{\cal C}_\ell}_B$ result is bang-on the Bayesian one. For
DMR53A+B, both the BC statistic and my modification give the same
value, lower than the Bayesian value with all modes included and
closer to the $k$-cut value described above: the two forms of
filtering are not that different. (Wright
\etal 1994 have also used the BC quadratic on the DMR data, and find
similar amplitudes.) Unfortunately, while well-suited to $Q_{rms,PS}$
estimation, these quadratic forms change with $n_{\Delta T}$, so joint
likelihoods should not be constructed, and, in any case, are a poor
way to get at $n_{\Delta T}$. The obvious first thing to try is
Weiner-filtering (with ${\cal E}_{TR,k}/(1+{\cal E}_{TR,k})$ acting on
$\xi_k$); however, not only does it change form with $n_{\Delta T}$,
it also changes form with $\sigma_{th}$, and so it is not useful for
either $Q_{rms,PS}$ or $n_{\Delta T}$ statistical analysis, although
it can be used to construct cleaned maps. I have also tried Bayesian
analyses without a residual, but with the mode sum in the
log-likelihoods suppressed by a measure $\propto {\cal
E}_{TR,k}^\alpha$: $\alpha =1$ gives results in agreement with the
BC-style quadratic statistics, but with smaller error bars because
it deals with the entire map.  Unfortunately, the result is not
stable to variations in $\alpha$, especially for FIRS, since lower
$\alpha$ does not sufficiently suppress the residual.

It is impractical to construct a likelihood when there are a number of
$\gamma^A$'s to compare with: to make smooth functions, too many
Monte Carlo runs are required even for this diagonal frame. Usually
some distance measure of observed from simulated $Q_A$ is used (\eg
the simple $chi^2$ measure for $C(\theta )$ used by the DMR team in
Smoot \etal (1992), and variants which include the effect of the strong
correlation from angle-bin to angle-bin in the theory, which I have
used Bond 1993).

What the $\gamma^A$ do is filter pixel-pairs, and this can have great
advantages over just linear filtering of modes.  For example, although
I do not know the details of the angular structure of the residual
noise, the pixel-error enhancement model contributes only to the zero
angle bin of the correlation function. If the residual's effect spills
over to neighbouring pixels, one can just cut out the core of the
correlation function, say out to $\theta_{fwhm}$. This would be an
excellent filter if some part of the residual arises because the beam
is not quite right or if there are white-noise or Poisson sources on
the sky. The same effective filtering does not happen with quadratic
band-power estimators. For example, white sky-noise gives a rising
$\ell^2$ contribution to ${\cal C}_\ell$. The residual in the FIRS map
obscured the interpretation of my results using quadratic band-power
estimators (which differ from the Hauser-Peebles form used by Wright
(1993) for DMR, which might be better).  Quadratic band-power
operators which exclude pixel-pairs out to $\theta_{fwhm}$ do filter
this rise, but the interpretation as power in an $\ell$-band is
lost. Thus, I believe that the Bayesian approach is superior to using
a quadratic set of band-power operators for ${\cal
C}_\ell$-estimation. And what comes out is an estimate of the
ensemble-averaged power, which is what we want, and not the specific
realization of the power that exists in a map, even if that map is our
own sky.

The filtering of small-angle systematic or physical effects in
correlation function analysis may make the $n_{\Delta T}$ determined
with $C(\theta )$ a better indication of the angular colour of the
large-angle sky than I get -- at least until the nature of the high
$k$-power is better understood.  For the FIRS map, there are
indications from $C(\theta )$ for a shallower slope, $n_{\Delta T}
\approx -1.7$ (Ganga and Page 1994); even so, it is clear from Fig.~2
that there are some anomalies in the $S/N$ band-powers that are not
favourable to $n_{\Delta T} < -2$.  Bennett \etal (1994) find
$n_{\Delta T} +3 = 1.59^{+0.49}_{-0.55}$ for two years of DMR data.
Modifications of the Bayesian approach to model a more extended
residual noise structure as well as the single-pixel level one is the
next step for me to better observe the angular colour of the CMB sky.

I would like to thank the FIRS and DMR teams for much discussion on how to
treat their maps, in particular Lyman Page, Al Kogut and George Smoot.
This research was supported by NSERC at
Toronto, CNRS in Paris, and the Canadian Institute for Advanced
Research.

\vspace{.2in}

\noindent
{\bf REFERENCES}
\vspace{.1in}

\begin{verse}
Bennett, C. \etal 1994, COBE prepint. \re
Bond, J. R. 1989, {\it The Formation of Cosmic Structure},
in {\it Frontiers in Physics --- From Colliders to
Cosmology}, p. 182-235, ed.
A. Astbury, B.A. Campbell, W. Israel, A.N. Kamal and F.C. Khanna,
Singapore: World Scientific.  \re
Bond, J. R. 1994, {\it
Cosmic Structure Formation and the Background Radiation},
in "Proceedings of the IUCAA Dedication Ceremonies," Pune,
India, Dec. 28-30, 1992, ed. T. Padmanabhan, Wiley \re
Bond, J.R.,  Crittenden, J.R.,
Davis, R.L., Efstathiou, G. and Steinhardt, P.J.,
1994a,  {\it Phys. Rev. Lett.} {\bf 72}, 13.
\re
Bond, J.R.,  Davis, R.L. and Steinhardt, P.J. 1994b, this volume.
\re
Bond, J. R. and Efstathiou, G.,
1987 {\it Mon.\ Not. R.\ astr.\ Soc.} {\bf226}, 655.
\re
Cheng, E.S., Cottingham, D.A., Fixsen, D.J., Inman, C.A., Kowitt,
M.S., Meyer, S.S., Page, L.A., Puchalla, J.L. and Silverberg, R.F.
1993, {\it Astrophys. J. Lett.} in press. [MSAM, {\it g2,g3}]
\re
Crittenden, R., Bond., J.R., Davis., R.L., Efstathiou, G. and Steinhardt, P.J.,
 1993, {\it Phys.\ Rev.\ Lett.}{\bf 71}, 324.
\re
de Bernardis, P. \etal 1993, this volume. [ARGO, {\it ar}]
\re
Dragovan, M. et al., 1993, Berkeley AAS  1993 Meeting. [{\it py}]
\re
Gaier, T., Schuster, J., Gunderson, J., Koch, T., Seiffert, M.,
Meinhold, P., and Lubin, P. 1992, {\it Astrophys.\ J.\ Lett.} {\bf 398} L1.
[{\it sp91}]\re
Ganga, K., Cheng, E., Meyer, S. and Page, L. 1993,  {\it Astrophys.\
J.\ Lett.} {\bf 410}, L57. [{\it firs}]
\re
Ganga, K.  and Page, L. 1994, in preparation. \re
Gunderson, J., {\it et \ al.}, 1993,  {\it Astrophys.\ J.\ Lett.} {\bf
413}, L1. [MAX, {\it M}]
\re
Kneissl, R. and Smoot, G. 1993, COBE note 5053
\re
Lineweaver, C. and Smoot, G. 1993, COBE note 5051
\re
Meinhold, P. and Lubin, P., 1991, {\it Astrophys.\
J.\ Lett.} {\bf 370}, 11. [{\it sp89}]
\re
Meinhold, P., Clapp, A., Cottingham, D., Devlin, M., Fischer, M.,
Gundersen, J., Holmes, W., Lange, A., Lubin, P., Richards, P. and
Smoot, G. 1993,
{\it Astrophys.\ J.\ Lett.} {\bf 409}, L1. [MAX, {\it M}]
\re
Readhead, A.C.S. {\it. et.\ al.}, 1989, {\it Astrophys.\ J.}\
{\bf 346}, 556. [{\it ov7}]
\re
Schuster, J., et al., 1993, {\it Astrophys.\ J.\ Lett.} {\bf 412}, L47.
[{\it sp91}]
\re
Smoot, G.F., et al., 1992 {\it Astrophys.\ J.\ Lett.} {\bf 396}, L45.
\re
Tucker, G.S.,   Griffin, G.S.,  Nguyen, H.  and  Peterson, J.B. 1993, .
{\it Astrophys.\ J.\ Lett.} {\bf 419}, L45. [{\it wd2, wd1}]
\re
Hancock, S., Davies, R., Lasenby, A.N., Guteirrez, C.M., Watson R.A.,
Rebolo, R. and Beckman, J.E. 1993, {\it Nature}, in press. [{\it ten}]
660. [{\it ten}]
\re
Wollack, E.J.,  Jarosik, N.C.,  Netterfield, C.B.,  Page, L.A., and
  Wilkinson, D.,  1993, Princeton preprint. [{\it bp}]\re
Wright, E. \etal 1994, {\it Astrophys.\ J.}, {\bf 420}, 1.\re
Wright, E. 1993, Annals N.Y. Acad. Sci. {\bf 668}, 836. \re
\end{verse}

\newpage

\begin{center}
{\bf FIGURE CAPTIONS}
\end{center}
\vspace{.2in}

\noindent
{\bf Figure 1.}  (a) and (b) give band-power estimates for the vintage
Fall 1993 data. The labels for the experiments are defined in the
text. The vertical error bars are 1 sigma, the inverted triangles are
90 or 95\% confidence limits.  Systematic errors or non-primary
sources (\eg dust, synchrotron) may contribute to these ${\cal
C}_\ell$'s; the underlying primary ${\cal C}_\ell$ may be lower, but
could be higher because of `destructive interference' among component
signals.  ${\cal C}_\ell$ is shown for a ``standard'' CDM model
($\Omega =1$, $n_s=1$, ${\rm h}=0.5$, $\Omega_B=0.05$) with normal
recombination to illustrate a typical Doppler peak, but the early
reionization (at $z> 200$) ${\cal C}_\ell$ is decidedly peak-less.
Whether a Doppler peak exists is unclear from the current data. (c)
Theoretical band-powers with (eventually) achievable 10\% error bars
are displayed for both the $n_s=1$ CDM model and a chaotic-inflation
inspired $n_s=0.95$ one -- including a gravity wave contribution
(Crittenden \etal 1993). The dotted ${\cal C}_\ell$ has
$\Omega_\Lambda \ne 0$ and high $H_0$ ($\Omega_\Lambda =0.75$,
$\Omega_{cdm}=0.22, \Omega_B=0.03, H_0=75$, $n_s=1$). This shows the
high precision we require to differentiate among models (for more
details, see Bond \etal 1994a,b).

\noindent
{\bf Figure 2.} $(S/N)$  band-powers ${\cal P}_{S/N, \beta}$
(eq.~\ref{eq:snbpow}) for the
orthogonal signal-to-noise eigenmodes of the FIRS and
DMR 53A+B maps, and for $n_{\Delta T}=-2,-1$ powers, as indicated.
The error bars denote observational
variance for the quadratics. The small open points denote
the $\beta$-band averages of  ${\cal
E}_{TR,k}$. The upper and lower curves about
these give the theoretical variances. All three curves scale with
$\sigma_{th}^2$. The triangles joined by the
dotted curve show the most likely residual noise the Bayesian analysis
finds. The observed ${\cal P}_{S/N, \beta}$
should be matched
within the errors by
$\sigma_{th}^2{\cal E}_{TR,k} + \sigma_{res}^2\overline{{\cal E}_{TR}}$
through adjustment of $\sigma_{th}$ and the residual
$\sigma_{res}$. For the DMR cases, the dashed line shows ${\cal
E}_{TR,k}$ for a $7^\circ$ Gaussian beam with no pixelization
correction. The data points also move a bit, but not enough for this
beam to explain the high $k$-behaviour. Small $k$ are, of course,
unaffected. Although $\sigma_{res}$ drops, $\sigma_{th}$ hardly does.

\noindent
{\bf Figure 3.} Contour maps (1-sigma to 10-sigma) of DMR and FIRS
likelihood functions for the scale invariant $n_{\Delta T}+3=1$
case. The heavy dot denotes the maximum.  Here, $\sigma_{th}=
[\avrg{{\cal C}_\ell }_{B}/10^{-10}]^{1/2}$ and the residual noise
amplitude $\sigma_{res}$ is normalized to have the same total power as
in the $\sigma_{th}$-band if $\sigma_{res}=\sigma_{th}$.  The 53A+B,
90A+B and FIRS maps show the same band-power detection, within the
errors, and in spite of the differing residuals. The 53 GHz difference
map, 53A-B, shows no large scale power, although the residual offset
remains at the same level as in A+B. The same holds for 90A-B.  As
$n_{\Delta T}$ increases, the required residual goes down for the
53A+B map, $\sigma_{res} \sim 0.466\sqrt{-n_{\Delta T}}$, with no need
of a residual for white noise ($n_{\Delta T}=0$), whereas it remains
constant at $\sigma_{res}\approx 2.5$ for the highly
inhomogeneously-sampled FIRS map, .

\end{document}